\documentclass[twocolumn]{aastex631}

\received{June 25, 2021}
\revised{August 12, 2021}
\accepted{August 23, 2021}

\shorttitle{VLA Survey of Late-time Radio Emission from SLSNe and the Host Galaxies}
\shortauthors{Hatsukade et al.}

\begin{document}

\title{A VLA Survey of Late-time Radio Emission from Superluminous Supernovae and the Host Galaxies}

\author[0000-0001-6469-8725]{Bunyo~Hatsukade}
\affiliation{Institute of Astronomy, Graduate School of Science, The University of Tokyo, 2-21-1 Osawa, Mitaka, Tokyo 181-0015, Japan}
\email{hatsukade@ioa.s.u-tokyo.ac.jp}

\author[0000-0001-8537-3153]{Nozomu~Tominaga}
\affiliation{National Astronomical Observatory of Japan, 2-21-1 Osawa, Mitaka, Tokyo 181-8588, Japan}
\affiliation{Department of Physics, Faculty of Science and Engineering, Konan University, 8-9-1 Okamoto, Kobe, Hyogo 658-8501, Japan}
\affiliation{Kavli Institute for the Physics and Mathematics of the Universe (WPI), The University of Tokyo, 5-1-5 Kashiwanoha, Kashiwa, Chiba 277-8583, Japan}

\author[0000-0001-7449-4814]{Tomoki~Morokuma}
\affiliation{Institute of Astronomy, Graduate School of Science, The University of Tokyo, 2-21-1 Osawa, Mitaka, Tokyo 181-0015, Japan}

\author[0000-0003-3932-0952]{Kana~Morokuma-Matsui}
\affiliation{Institute of Astronomy, Graduate School of Science, The University of Tokyo, 2-21-1 Osawa, Mitaka, Tokyo 181-0015, Japan}

\author[0000-0003-1747-2891]{Yuichi~Matsuda}
\affiliation{National Astronomical Observatory of Japan, 2-21-1 Osawa, Mitaka, Tokyo 181-8588, Japan}
\affiliation{Graduate University for Advanced Studies (SOKENDAI), Osawa 2-21-1, Mitaka, Tokyo 181-8588, Japan}

\author[0000-0003-4807-8117]{Yoichi~Tamura}
\affiliation{Department of Physics, Nagoya University, Furo-cho, Chikusa-ku, Nagoya 464-8602, Japan}

\author[0000-0002-8169-3579]{Kotaro~Niinuma}
\affiliation{Graduate School of Sciences and Technology for Innovation, Yamaguchi University, Yoshida 1677-1, Yamaguchi, Yamaguchi 753-8512, Japan}

\author[0000-0002-3789-770X]{Kazuhiro~Motogi}
\affiliation{Graduate School of Sciences and Technology for Innovation, Yamaguchi University, Yoshida 1677-1, Yamaguchi, Yamaguchi 753-8512, Japan}

\begin{abstract}

We present the results of 3 GHz radio continuum observations of 23 superluminous supernovae (SLSNe) and their host galaxies by using the Karl G. Jansky Very Large Array conducted 5--21 years after the explosions. 
The sample consists of 15 Type I and 8 Type II SLSNe at $z < 0.3$, providing one of the largest sample of SLSNe with late-time radio data. 
We detected radio emission from one SLSN (PTF10hgi) and 5 hosts with a significance of $>$5$\sigma$. 
No time variability is found in late-time radio light curves of the radio-detected sources in a timescale of years except for PTF10hgi, whose variability is reported in a separate study. 
Comparison of star-formation rates (SFRs) derived from the 3 GHz flux densities with those derived from SED modeling based on UV--NIR data shows that four hosts have an excess of radio SFRs, suggesting obscured star formation. 
Upper limits for undetected hosts and stacked results show that the majority of the SLSN hosts do not have a significant obscured star formation. 
By using the 3 GHz upper limits, we constrain the parameters for afterglows arising from interaction between initially off-axis jets and circumstellar medium (CSM).  
We found that the models with higher energies ($E_{\rm iso} \gtrsim$ several $\times$ $10^{53}$ erg) and CSM densities ($n \gtrsim 0.01$ cm$^{-3}$) are excluded, but lower energies or CSM densities are not excluded with the current data. 
We also constrained the models of pulsar wind nebulae powered by a newly born magnetar for a subsample of SLSNe with model predictions in the literature. 

\end{abstract}

\keywords{
\href{http://astrothesaurus.org/uat/2008}{Radio transient sources (2008)}; 
\href{http://astrothesaurus.org/uat/508}{Extragalactic radio sources (508)}; 
\href{http://astrothesaurus.org/uat/1668}{Supernovae (1668)}; 
\href{http://astrothesaurus.org/uat/241}{Circumstellar matter (241)};
\href{http://astrothesaurus.org/uat/1340}{Radio continuum emission (1340)}; 
\href{http://astrothesaurus.org/uat/1766}{Very Large Array (1766)}; 
\href{http://astrothesaurus.org/uat/1338}{Radio Astronomy (1338)}
}

\section{Introduction}

Superluminous supernovae (SLSNe) are very bright explosions ($>$10--100 times brighter) and rare events ($<$100 times lower) compared to ordinary SNe \citep[][for reviews]{gal-12, gal-19, mori18a}. 
SLSNe are detected at high redshifts up to $z \sim 4$ \citep{cook12, mori19a}, and therefore can be powerful probes of the distant universe. 
The power source and mechanism for large luminosities are still a matter of debate. 
SLSNe are classified into two types according to the spectra: 
hydrogen-poor Type I SLSNe (or SLSNe-I) and hydrogen-rich Type II SLSNe (or SLSNe-II). 
SLSNe-II can be explained as an interaction between the SN ejecta and dense circumstellar medium \citep[CSM; e.g.,][]{woos07}, while a number of models have been proposed for SLSNe-I: 
a large amount of $^{56}$Ni produced by a pair-instability SN \citep[e.g.,][]{woos07}, 
spin-down of a newborn strongly magnetized neutron star \citep[magnetar; e.g.,][]{kase10, woos10}, 
fallback accretion onto a compact remnant \citep[e.g.,][]{dext13}, 
and interaction with dense CSM \citep[e.g.,][]{smit07a, chev11}.

Radio observations can probe synchrotron emission arising from shock interaction of SN ejecta or jet with surrounding material, providing useful constraints on the models of SLSNe. 
It is expected that late-time radio emission may be caused by initially off-axis jets that decelerate and spread into the line of sight. 
It is also predicted that quasistate synchrotron emission may arise from pulsar wind nebulae (PWNe) powered by a newly born magnetar \citep{metz14, mura16, kash17, metz17}. 
The radio emission is initially absorbed in the PWNe and SN ejecta, but the system can be transparent on timescales of decades. 
These central engine models are also a possible model for long-duration gamma-ray bursts (LGRBs) \citep[e.g.,][]{metz15, marg18b}, suggesting a connection between SLSNe-I and LGRBs. 
It is interesting that the magnetar model is also one of the plausible models for fast radio bursts (FRBs), which are mysterious radio transients with millisecond-scale bright flashes \citep[][for a review]{cord19}. 
Magnetar models have been applied to the origin of FRB\,121102 \citep[e.g.,][]{mura16, kash17, metz17, marg18a, marg18}. 
Recently, FRB\,200428 was identified as a Galactic magnetar, SGR~$1935+2154$ \citep{ande20, boch20a}, showing a magnetar origin of at least one FRB.

Previous radio observations of SLSNe resulted in nondetections in most cases, providing constraints on physical properties (such as energies, mass-loss rates, and CSM densities) and models for SLSNe-I \citep[e.g.,][]{nich16a, nich18, marg18b, bose18, copp18, law19, efte21}. 
\cite{efte19} found an unresolved radio source coincident with the position of SLSN-I PTF10hgi. 
They argued that the radio emission is consistent with an off-axis jet or wind nebula powered by a magnetar born in the SLSN, suggesting the presence of a central engine. 
\cite{law19} and \cite{mond20} also supported the model of magnetar-powered SLSN based on the observations at 0.6--15 GHz. 
\cite{hats21} found a variability in late-time radio emission in PTF10hgi for the first time among SLSNe. 
They constrained both the rise and decay phases of the radio light curve over three years, peaking at approximately 8--9 yr after the explosion. 
They concluded that plausible scenarios are a low-luminosity active galactic nucleus (AGN) in the host galaxy or a magnetar wind nebula by considering the radio light curve and spectrum.

In order to constrain the models, it is also important to understand the properties of their host galaxies. 
Previous studies have shown that SLSN-I hosts are typically dwarf galaxies with low-luminosity, low stellar mass, low star-formation rate (SFR), and high specific SFR (sSFR) compared to local star-forming galaxies and the hosts of core-collapse SNe, while SLSN-II hosts show a wider range of those parameters \citep[e.g.,][]{lunn14, lelo15, angu16, perl16, chen17, schu18, tagg21}. 
The observations of SLSN hosts have been conducted mainly in the optical/near-infrared (NIR) wavelengths, which are subject to dust extinction in contrast to longer wavelengths, and it is possible that we are missing dust-obscured star formation. 
Radio observations are important to probe the properties of star formation in the SLSN host without the effect of dust extinction \citep{schu18, hats18, arab19a, hats20, efte21}. 
\cite{schu18} searched radio emission for a sample of SLSN hosts by using the survey data of Faint Images of the Radio Sky at Twenty-Centimeters \citep[FIRST;][]{beck95}, the NRAO VLA Sky Survey \citep[NVSS;][]{cond98}, and the Sydney University Molonglo Sky Survey \citep[SUMSS;][]{bock99}, and did not find radio detection. 
They also obtained upper limits on three SLSN hosts based on deep 1.5 GHz observations with the Karl G. Jansky Very Large Array (VLA). 
\cite{hats18} performed 3 GHz radio observations of the host galaxies of 8 SLSNe (5 Type I and 3 Type II) at $0.1 < z < 0.3$ with VLA. 
They found the excess of radio-based SFRs over extinction-corrected optically based SFRs (UV-based SFRs from SED fitting or H$\alpha$-based SFRs) in three hosts, suggesting the existence of dust-obscured star formation that were missed in previous observations. 
This indicates the necessity of longer-wavelength observations in order to understand true star-forming activity in SLSN hosts. 
They also found that three hosts, which were located within the range of the main sequence based on previous optical observations, are actually above the main sequence in our radio observations, suggesting that they have a starburst nature. 
\cite{efte21} observed 15 SLSNe-I at 6 GHz with VLA and 29 SLSNe-I at 100 GHz with the Atacama Large Millimeter/submillimeter Array (ALMA), seven of which overlapped. 
They did not detect emission from any of the SLSNe except for a 6 GHz detection of PTF10hgi. 
Only the host of PTF12dam was detected at 6 GHz, concluding that there is no significant dust obscuration in the SLSNe hosts.

In this paper, we present the results of VLA 3 GHz continuum observations of 23 SLSNe (15 Type I and 8 Type II) and their host galaxies, offering one of the largest sample of SLSNe with radio data. 
The remainder of the paper is organized as follows. 
Section~\ref{sec:observations} describes targets, VLA observations, and data reduction. 
The results are shown in Section~\ref{sec:results}. 
In Section~\ref{sec:discussion}, we discuss a magnetar model for the SLSNe and obscured star formation in the host galaxies. 
Conclusions are presented in Section~\ref{sec:conclusions}. 
Throughout the paper, we adopt a \cite{chab03} initial mass function and cosmological parameters of $H_0=67.7$ km s$^{-1}$ Mpc$^{-1}$ and $\Omega_{\rm{M}}=0.310$ based on the {\sl Planck} 2018 results \citep{plan20f}.

\begin{deluxetable*}{lcclCR
}
\tablecaption{Targets \label{tab:targets}}
\tablewidth{0pt}
\tablehead{
\colhead{SLSN} & \colhead{Discovery Date\tablenotemark{a}} & \colhead{Type} & \colhead{$z$}
& \colhead{$\log{\rm SFR}$\tablenotemark{b}} & \colhead{$\log{M_{*}}$\tablenotemark{b}} \\
 & & & & \colhead{($M_{\odot}$~yr$^{-1}$)} & \colhead{($M_{\odot}$)} 
}
\startdata
CSS121015\tablenotemark{c} & 2012-10-15 & II & 0.286  & -0.52 ^{+0.38 }_{-0.29 } &  8.15^{+0.15}_{-0.17} \\
LSQ12dlf                   & 2012-07-10 & I  & 0.255  & -1.36 ^{+0.54 }_{-0.43 } &  7.56^{+0.33}_{-0.34} \\
LSQ14mo                    & 2014-01-30 & I  & 0.2561 & -0.84 ^{+0.42 }_{-0.34 } &  7.89^{+0.15}_{-0.19} \\
PTF09cnd                   & 2009-08-07 & I  & 0.258  & -0.64 ^{+0.21 }_{-0.18 } &  7.87^{+0.20}_{-0.21} \\
PTF10hgi\tablenotemark{d}  & 2010-05-15 & I  & 0.0987 & -1.02 ^{+0.44 }_{-0.52 } &  7.58^{+0.29}_{-0.31} \\
PTF10qaf                   & 2010-08-05 & IIn& 0.284  &  0.11 ^{+0.50 }_{-0.50 } &  8.73^{+0.22}_{-0.25} \\
PTF10uhf                   & 2010-08-25 & I  & 0.2882 &  0.834^{+0.122}_{-0.263} & 11.23^{+0.15}_{-0.12} \\
PTF12dam                   & 2012-04-10 & I  & 0.107  & -0.00 ^{+0.27 }_{-0.26 } &  8.89^{+0.15}_{-0.30} \\
SN\,1999as                 & 1999-02-18 & I  & 0.127  &  0.14 ^{+0.37 }_{-0.35 } &  8.94^{+0.20}_{-0.17} \\
SN\,1999bd                 & 1999-02-19 & IIn& 0.151  &  0.037^{+0.118}_{-0.162} &  9.10^{+0.22}_{-0.11} \\
SN\,2006gy                 & 2006-09-18 & IIn& 0.019  & -1.12 ^{+0.08 }_{-0.08 } & 11.70^{+0.06}_{-0.21} \\
SN\,2006tf\tablenotemark{e}& 2006-12-12 & IIn& 0.074  & -1.25 ^{+0.48 }_{-0.37 } &  7.54^{+0.47}_{-0.20} \\
SN\,2007bi\tablenotemark{f}& 2007-04-06 & I  & 0.128  & -1.71 ^{+0.53 }_{-0.52 } &  7.92^{+0.20}_{-0.21} \\
SN\,2007bw\tablenotemark{g}& 2007-04-18 & IIn& 0.140  & -0.24 ^{+0.47 }_{-0.37 } &  9.39^{+0.19}_{-0.09} \\
SN\,2008es\tablenotemark{h}& 2008-04-26 & II & 0.205  & -1.99 ^{+0.28 }_{-0.27 } &  6.19^{+0.33}_{-0.36} \\
SN\,2009nm\tablenotemark{i}& 2009-11-20 & IIn& 0.210  & -0.60 ^{+0.65 }_{-0.62 } &  8.65^{+0.33}_{-0.34} \\
SN\,2010kd                 & 2010-11-14 & I  & 0.101  & -0.98 ^{+0.44 }_{-0.31 } &  7.30^{+0.25}_{-0.29} \\
SN\,2011ep\tablenotemark{j}& 2011-04-14 & I  & 0.280  &  0.05 ^{+0.41 }_{-0.30 } &  7.79^{+0.42}_{-0.36} \\
SN\,2011ke\tablenotemark{k}& 2011-04-06 & I  & 0.143  & -0.82 ^{+0.24 }_{-0.23 } &  7.50^{+0.20}_{-0.18} \\
SN\,2011kf\tablenotemark{l}& 2011-12-30 & I  & 0.245  & -0.86 ^{+0.18 }_{-0.20 } &  7.58^{+0.19}_{-0.22} \\
SN\,2012il\tablenotemark{m}& 2012-01-19 & I  & 0.175  & -0.74 ^{+0.22 }_{-0.36 } &  8.20^{+0.18}_{-0.17} \\
SN\,2013dg\tablenotemark{n}& 2013-05-17 & I  & 0.265  & -1.43 ^{+0.80 }_{-0.52 } &  7.09^{+0.82}_{-0.70} \\
SN\,2015bn\tablenotemark{o}& 2014-12-23 & I  & 0.110  & -1.06 ^{+0.69 }_{-0.50 } &  7.50^{+0.38}_{-0.35} \\
\enddata
\tablenotetext{a}{Discovery date (maximum date for PTF10qaf) taken from Open Supernova Catalog\footnote{https://sne.space/} \citep{guil17}.} 
\tablenotetext{b}{SFR and stellar mass of host galaxies derived from SED modeling by \cite{schu18}.} 
\tablenotetext{c}{Alias: CSS121015:004244$+$132827}
\tablenotetext{d}{Aliases: SN\,2010md, PSO J249.4461$+$06.2081.}
\tablenotetext{e}{Alias: CSS070320:124616+112555.}
\tablenotetext{f}{Alias: SNF20070406$-$008.}
\tablenotetext{g}{Alias: SNF20070418$-$020.}
\tablenotetext{h}{Alias: ROTSE3 J115649.1$+$542726.}
\tablenotetext{i}{Alias: CSS091120:100525$+$511639.}
\tablenotetext{j}{Alias: CSS110414:170342$+$324553.}
\tablenotetext{k}{Aliases: CSS110406:135058+261642, PS1-11xk, PTF11dij.}
\tablenotetext{l}{Alias: CSS111230:143658$+$163057.}
\tablenotetext{m}{Aliases: CSS120121:094613$+$195028, PS1-12fo.}
\tablenotetext{n}{Aliases: CSS130530:131841$-$070443, MLS130517:131841$-$070443.}
\tablenotetext{o}{Aliases: CSS141223:113342$+$004332, MLS150211:113342$+$004333, PS15ae.}
\end{deluxetable*}

\section{VLA Observations} \label{sec:observations}

\subsection{Targets} \label{sec:targets}

In order to understand the general properties of SLSNe and their hosts, it is essential to conduct an unbiased search for radio emission by using a complete sample. 
The targets of VLA observations were selected from the list of SLSNe compiled by \cite{schu18}. 
They made a list of 69 SLSNe discovered before the end of 2014 among all SLSNe reported in the literature, providing the largest and complete catalog of SLSNe. 
The redshifts of the sample are $z = 0.1$--2 with a singular object at $z = 3.899$, and the median redshifts are $z = 0.46$ and 0.21 for SLSNe-I and SLSNe-II, respectively. 
From the list, we selected SLSNe well observable with VLA (declination $> -25$ deg) and located at $z < 0.3$ to ensure significant constraint on obscured star formation. 
In order to avoid the contamination from AGN to radio emission, we excluded CSS100217, whose host is known to have possible AGN features \citep{lelo15, perl16, schu18}. 
We also excluded six SLSNe (PTF10aagc, PTF11rks, SN2005ap, SN2008am, SN2008fz, and SN2010gx) that already have deep radio upper limits as reported in \cite{hats18} and \cite{schu18}. 
This results in 23 targets that consists of 15 Type I and 8 Type II SLSNe (Table~\ref{tab:targets}). 
This is one of the largest samples of SLSNe with radio observations. 
All of them were observed in the VLA 1.4 GHz surveys of FIRST and NVSS, and none was detected in relatively shallow observations \citep[$\sim$0.15 and $\sim$0.45 mJy~beam$^{-1}$, respectively;][]{schu18}. 
In order to examine the time variability of the radio emission, four SLSNe (SN1999bd, PTF10qaf, PTF10uhf, PTF12dam) with radio detection by \cite{hats18} were included.
Four SLSNe (SN1999bd, PTF10qaf, PTF10uhf, PTF12dam) whose hosts were detected in previous observations \citep{hats18} were included in the targets in order to examine the time variability of the radio emission. 
Our observations of these radio-detected sources were conducted $\gtrsim$2 yr after the first observations, allowing us to examine the time variability. 
Detections of radio emission from PTF10hgi were reported by \cite{efte19}, \cite{law19}, and \cite{mond20} after the target selection of this work. 
Our VLA observations of PTF10hgi were reported in detail in a separate paper \citep{hats21}.

We note previous radio observations of the targets at $\sim$1--10 GHz. 
Existing radio observations of SLSNe and their hosts are compiled by \cite{copp18}, \cite{schu18}, and \cite{efte21}. 
Of our targets, PTF09cnd, SN\,2012il, and SN\,2015bn were observed within a year after the explosions, yielding nondetections \citep{chan09, chan10, chom12, nich16a, alex16}. 
Late-time radio observations ($\gtrsim$a few years after the explosions) by \cite{law19} and \cite{efte21} covered a fraction of the SLSNe-I in our sample; 
PTF09cnd, PTF10hgi, SN\,2007bi, SN\,2010kd, and SN\,2011ke at 3 GHz by \cite{law19}; 
LSQ12dlf, PTF09cnd, PTF12dam, SN\,1999as, SN\,2007bi, SN\,2010kd, SN\,2011ke, SN\,2011kf, SN\,2012il at 6 GHz by \cite{efte21}.

\begin{deluxetable*}{lcccclRRcC}
\tablecaption{VLA 3 GHz Observations \label{tab:observations}}
\tablewidth{0pt}
\tablehead{
\colhead{SLSN} & \colhead{Obs. Date} & \colhead{Time Since Expl.\tablenotemark{a}} & \colhead{$T_{\rm on}$} 
& \colhead{$N_{\rm ant}$} & \colhead{Baseline} & \colhead{Beam Size} & \colhead{P.A.} 
& \colhead{RMS} & \colhead{$S_{\rm 3 GHz}$} \\ 
\colhead{} & \colhead{} & \colhead{(yr)} & \colhead{(min)} 
& \colhead{} & \colhead{(m)} & \colhead{($''$)} & \colhead{($^{\circ}$)} 
& \colhead{($\mu$Jy beam$^{-1}$)} & \colhead{($\mu$Jy)} 
}
\decimalcolnumbers
\startdata
CSS121015  & 2019-01-02 &  4.8 &  91 & 27 & 45--3400 &  8.4 \times 5.4 & -1.6  & 5.7 & <17      \\
LSQ12dlf   & 2019-01-03 &  5.2 &  91 & 26 & 45--3400 & 13.0 \times 5.2 &   4.7 & 6.4 & <19      \\
...        & 2020-04-01 &  6.2 & 103 & 27 & 45--3400 & 11.7 \times 5.7 &  12.6 & 6.1 & <18      \\
...        & combined   & ...  & ... & ...& ...      & 12.1 \times 5.5 &   8.4 & 5.3 & <16      \\
LSQ14mo    & 2020-04-19 &  4.9 & 103 & 27 & 45--3400 & 10.4 \times 5.7 & -11.2 & 7.0 & <21      \\
PTF09cnd   & 2018-11-27 &  7.4 &  91 & 24 & 78--3400 &  8.2 \times 4.9 & -15.8 & 5.4 & <16      \\
PTF10hgi   & 2018-12-02 &  7.8 &  91 & 26 & 45--3400 &  8.9 \times 5.8 &  42.7 & 5.3 & 85 \pm 7\tablenotemark{b} \\
...        & 2020-04-25 &  9.1 & 103 & 28 & 45--3400 &  7.8 \times 6.3 &  43.5 & 5.8 & 51 \pm 6\tablenotemark{b} \\
PTF10qaf   & 2017-05-29 &  1.9 &  91 & 27 & 45--3400 &  8.1 \times 6.1 & -49.1 & 5.4 & 58 \pm 6\tablenotemark{c} \\
...        & 2020-04-06 &  4.1 & 103 & 27 & 45--3400 &  8.1 \times 6.1 &  49.6 & 6.9 & 57 \pm 8\tablenotemark{c} \\
...        & combined   & ...  & ... & ...& ...      &  6.9 \times 6.7 &  86.4 & 4.9 & 57 \pm 6\tablenotemark{c} \\
PTF10uhf   & 2017-05-29 &  5.2 &  91 & 27 & 45--3400 &  6.7 \times 5.5 &  73.5 & 5.4 & 82 \pm 7\tablenotemark{d} \\
...        & 2020-06-08 &  7.6 & 103 & 27 & 45--3400 &  7.7 \times 5.7 &  89.7 & 5.4 & 77 \pm 7\tablenotemark{d} \\
...        & combined   & ...  & ... & ...& ...      &  7.0 \times 5.6 &  82.8 & 4.0 & 77 \pm 6\tablenotemark{d} \\
PTF12dam   & 2017-05-28 &  4.6 &  91 & 27 & 45--3400 &  6.5 \times 5.6 &  68.8 & 4.5 & 139\pm 8\tablenotemark{e} \\
...        & 2020-06-06 &  7.4 & 103 & 26 & 45--3200 &  7.0 \times 5.4 &  67.2 & 4.6 & 131\pm 8\tablenotemark{e} \\
...        & combined   & ...  & ... & ...& ...      &  6.5 \times 5.6 &  82.8 & 3.5 & 133\pm 8\tablenotemark{e} \\
SN\,1999as & 2019-01-14 & 17.7 &  91 & 24 & 45--3400 &  6.9 \times 5.8 & -50.3 & 5.5 & <17      \\
SN\,1999bd & 2017-05-28 & 15.9 &  91 & 27 & 45--3400 &  6.2 \times 5.3 & -2.8  & 5.2 & 43 \pm 6\tablenotemark{f} \\
...        & 2020-04-14 & 18.4 & 103 & 26 & 45--3400 &  6.2 \times 5.6 &  22.4 & 5.4 & 37 \pm 6\tablenotemark{f} \\
...        & combined   & ...  & ... & ...& ...      &  6.0 \times 5.4 & -11.7 & 4.0 & 41 \pm 5\tablenotemark{f} \\
SN\,2006gy & 2019-01-01 & 12.1 &  91 & 25 & 45--3400 &  6.2 \times 5.3 & -63.0 & 44  & 838\pm61\tablenotemark{g} \\
...        & 2020-03-16 & 13.2 & 103 & 27 & 45--3400 &  5.8 \times 5.2 &  74.7 & 43  & 859\pm61\tablenotemark{g} \\
...        & combined   & ...  & ... & ...& ...      &  6.0 \times 5.3 & -69.8 & 38  & 859\pm57\tablenotemark{g} \\
SN\,2006tf & 2020-06-04 & 12.5 & 103 & 26 & 45--3400 &  6.7 \times 5.8 & -0.8  & 5.0 & <15      \\
SN\,2007bi & 2020-05-22 & 11.6 & 103 & 27 & 45--3400 &  8.8 \times 5.9 &  52.5 & 7.1 & <21      \\
SN\,2007bw & 2019-02-19 & 10.4 &  91 & 27 & 45--11100&  3.2 \times 2.2 &  14.9 & 4.6 & <14 \\
SN\,2008es & 2019-02-17 &  9.0 &  91 & 27 & 45--11100&  3.6 \times 1.9 &  49.2 & 4.2 & <12      \\
SN\,2009nm & 2019-02-19 &  7.6 &  91 & 27 & 45--11100&  3.4 \times 1.8 &  21.6 & 4.1 & <12      \\
SN\,2010kd & 2019-02-19 &  7.5 &  91 & 27 & 45--11100&  3.3 \times 1.8 &  21.4 & 3.9 & <12      \\
SN\,2011ep & 2019-02-22 &  6.1 &  91 & 26 & 80--11100&  3.4 \times 2.2 &  42.2 & 4.3 & <13      \\
SN\,2011ke & 2020-04-24 &  7.9 & 109 & 28 & 45--3400 &  6.7 \times 6.4 &  69.6 & 4.3 & <13      \\
SN\,2011kf & 2020-06-05 &  6.8 & 103 & 27 & 45--3400 &  9.8 \times 5.8 & -58.4 & 6.1 & <18      \\
SN\,2012il & 2019-02-22 &  6.0 &  91 & 26 & 80--11100&  3.4 \times 1.7 &   2.3 & 4.0 & <12      \\
SN\,2013dg & 2020-06-06 &  5.6 & 103 & 27 & 45--3400 &  8.9 \times 6.0 & -8.3  & 24  & <72      \\
SN\,2015bn & 2020-06-07 &  4.9 & 103 & 27 & 45--3400 & 11.2 \times 6.5 &  51.9 & 15  & <45      \\
\enddata
\tablecomments{
(1) SLSN name. 
(2) VLA observing date. ``combined'' is for the combined results of two observations. 
(3) Time since explosion (rest frame). 
(4) On-source integration time. 
(5) Number of antennas. 
(6) Range of baseline lengths. 
(7) Synthesized beam size. 
(8) Position angle of synthesized beam. 
(9) RMS noise level of the map.  
(10) Flux density at 3 GHz. Limits are 3$\sigma$. 
\tablenotetext{a}{Explosion rest frame.}
\tablenotetext{b}{Likely to be from both SN and host; \citet{hats21}}
\tablenotetext{c}{Likely to be from host.}
\tablenotetext{d}{Integrated intensity measured with the CASA {\tt\string imfit} task. Likely to be from host.}
\tablenotetext{e}{Likely to be from host (SDSS\,J142446.21$+$461348.6).}
\tablenotetext{f}{Likely to be from host (A093029$+$1626).}
\tablenotetext{g}{Likely to be from host (NGC 1260).}
}
\end{deluxetable*}

\subsection{Observations and Data Reduction} \label{sec:reduction}

The VLA S-band 3 GHz (13 cm) observations were performed from November 2018 to February 2019 in semester 18B (Project ID: 18B-077) and from April to June 2020 in semester 20A (Project ID: 20A-133). 
The observing dates are 5--21 yr after the discovery of the SLSNe. 
Observations were done in the array configuration C except for six targets (SN2007bw, SN2008es, SN2009nm, SN2010kd, SN2011ep, and SN2012il) that were observed during a transition period from configuration C to B. 
The number of antennae used in the observations was 24--28. 
The WIDAR correlator was used with 8-bit samplers. 
We used two basebands with 1 GHz bandwidth centered at 2.5 GHz and 3.5 GHz, which provided a total bandwidth of 2 GHz. 
The field of view was $7\farcm4$ (full width at half power). 
The positions of the SNe were used as phase centers. 
Bandpass and amplitude calibrations were conducted with 3C147 or 3C286, and phase calibrations were conducted with nearby quasars. 
The total observing time of each target was 91--109 min. 
Details of the observations appear in Table~\ref{tab:observations}.

The data were reduced with Common Astronomy Software Applications \citep[CASA;][]{mcmu07} release 5.6.2. 
About 10\%--15\% of the data were flagged by the pipeline processing. 
The maps were produced with the task {\tt\string tclean}. 
The Briggs weighting with {\tt\string robust} 0.5 was adopted. 
The absolute flux accuracy was estimated by comparing the measured flux density of the amplitude calibrators and the flux density scale of \cite{perl17}, and the difference was found to be $<$2\%. 
For detected sources, we conducted a 2D Gaussian fit to the radio emission in the image plane. 
PTF10uhf was spatially resolved, and we measured an integrated flux density by using the {\tt\string imfit} task. 
The others were not spatially resolved or only marginally resolved and we adopted the peak intensity as a source flux density. 
The flux uncertainty reported here is a combination of the map rms and a 5\% absolute flux calibration uncertainty\footnote{\url{https://science.nrao.edu/facilities/vla/docs/manuals/oss/performance/fdscale}}.

We also reanalyzed the VLA data of \cite{hats18} of four SLSNe (PTF10qaf, PTF10uhf, PTF12dam, and SN\,1999bd) that were included in the sample of this study by using the same version of CASA as used in this work. 
Although \cite{hats18} reported that PTF10qaf is spatially resolved, we found that the emission is only marginally resolved, and we adopted a peak intensity as a source flux density in this work. 
In order to create deeper images, we combined the data of \cite{hats18} with the newly obtained data except for PTF10hgi, which showed a time variability. 
The results are included in Table~\ref{tab:observations}.

\section{Results} \label{sec:results}

\begin{figure*}
\begin{center}
\includegraphics[width=.9\linewidth]{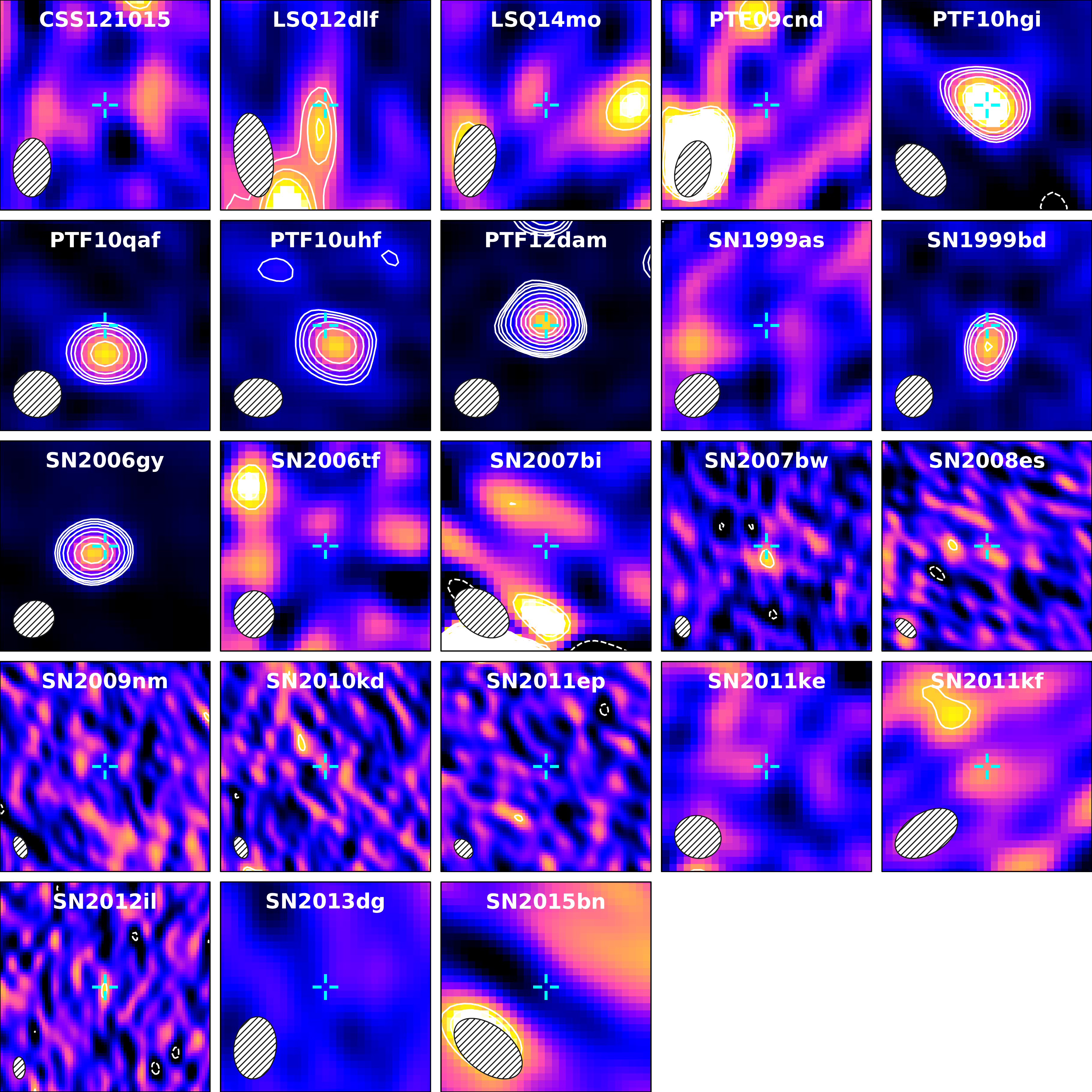}
\end{center}
\caption{
VLA 3 GHz continuum maps centered at SN positions. 
Image size is $30'' \times 30''$. 
North is up, and east is to left. 
Crosses represent SN positions.
Contours are $-3\sigma$, $3\sigma$, $4\sigma$, $5\sigma$, $7\sigma$, $10\sigma$, and $5\sigma$ steps subsequently (negative contours as dashed). 
Synthesized beam size is shown at lower left corners. 
Image of SN\,2006gy is based on combined data of two observations in this study, and images of PTF10qaf, PTF10uhf, PTF12dam, and SN\,1999bd are based on combined data of this study and \cite{hats18}. 
Image of PTF10hgi is based on data taken in 2018. 
\label{fig:radio}}
\end{figure*}

\begin{figure}
\begin{center}
\includegraphics[width=\linewidth]{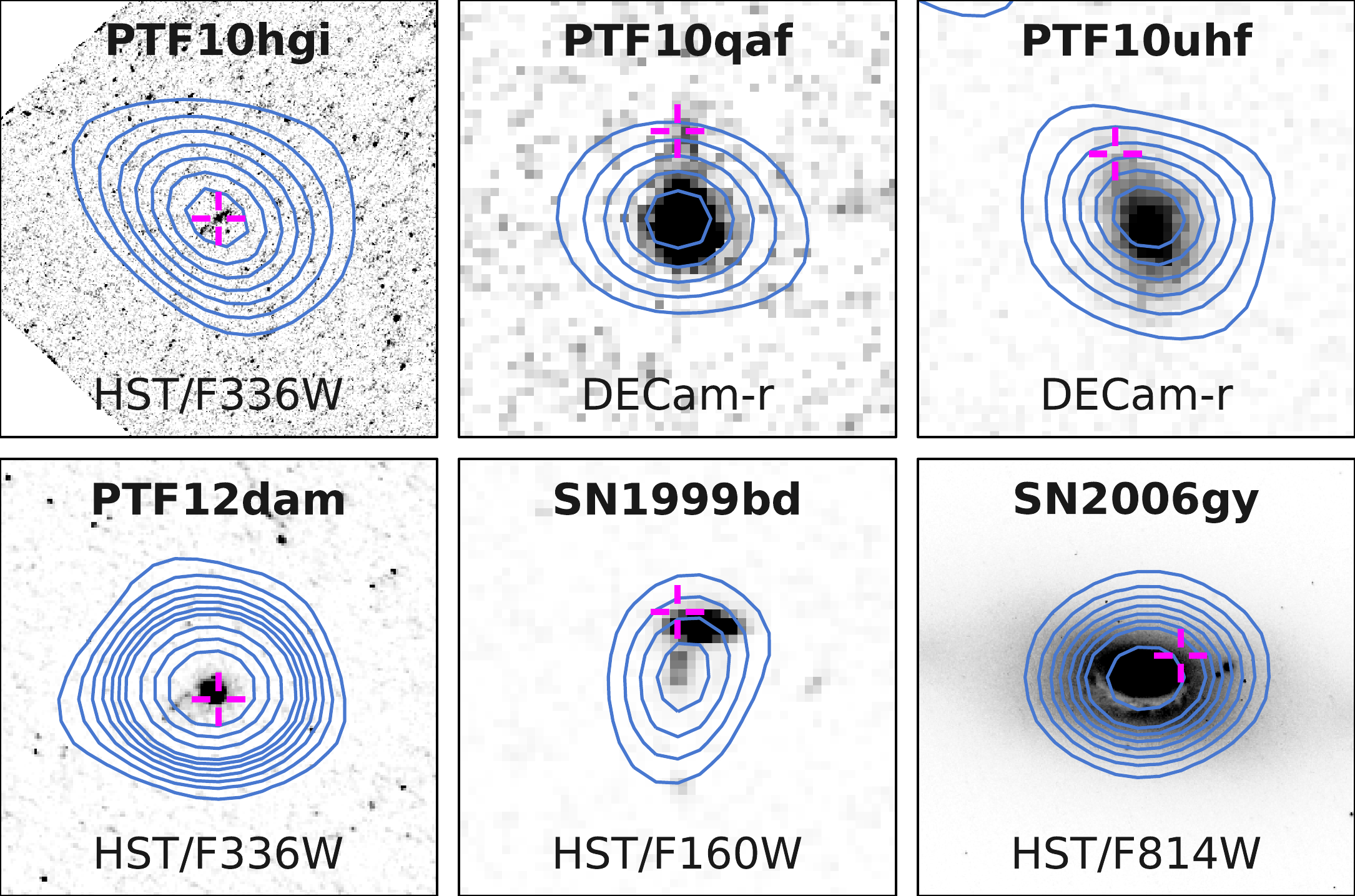}
\end{center}
\caption{
VLA 3 GHz contours overlaid on optical/NIR images for radio-detected ($>$5$\sigma$) sources centered at radio peak positions. 
Image size is $20'' \times 20''$. 
North is up, and east is to left. 
Crosses represent SN positions.
Contour levels are same as in Figure~\ref{fig:radio}. 
\label{fig:optical}}
\end{figure}

The VLA 3 GHz continuum maps are shown in Figure~\ref{fig:radio}. 
We detected radio emission in six sources (PTF10hgi, PTF10qaf, PTF10uhf, PTF12dam, SN\,1999bd, SN\,2006gy) with a peak signal-to-noise ratio (S/N) above 5, and tentatively detected in two sources (SN\,2007bw and SN\,2012il) with S/N $\sim$ 3.6. 
Five of them (PTF10hgi, PTF10qaf, PTF10uhf, PTF12dam, and SN\,1999bd) were detected in the previous 3 and 6 GHz observations \citep{hats18, law19, efte21}, and the peak positions are consistent. 
SN\,2006gy and PTF10hgi were detected in both semesters (18B and 20A) in our observations. 
Because the radio emission detected in the map of LSQ12dlf was $\sim$3$''$ away from the SN position, which is larger than the positional uncertainty of the radio observations ($\sim$2$''$), we consider that the emission is not associated with the SN or its host galaxy (Table~\ref{tab:offset}). 
Figure~\ref{fig:optical} shows the radio contours for the six radio-detected sources overlaid on the optical/NIR images taken from the Hubble Legacy Archive\footnote{\url{https://hla.stsci.edu/}} and the public data of the Dark Energy Survey \citep[DES;][]{abbo18, morg18, flau15}. 
The radio emission of PTF10qaf, PTF10uhf, and SN\,2006gy are dominated by the galaxy center rather than the SN positions. 
The radio peak position of PTF12dam is also slightly offset from the SN position and close to the galaxy center. 
This is supported by the higher angular resolution ($0\farcs95 \times 0\farcs74$) observations at 6 GHz by \cite{efte21}, who claimed that the emission was most likely related to star formation in the host. 
The peak radio position of SN\,1999bd appears to be associated with a faint object $\sim$2$''$ south of the SN position. 
While the apparent distance between the radio peak and the SN position is still within the uncertainty of the radio observations, it is not clear whether the emission comes from the SN or its host galaxy. 
It is possible that the radio emission arose from an unassociated foreground/background galaxy.

\begin{figure}
\begin{center}
\includegraphics[width=.9\linewidth]{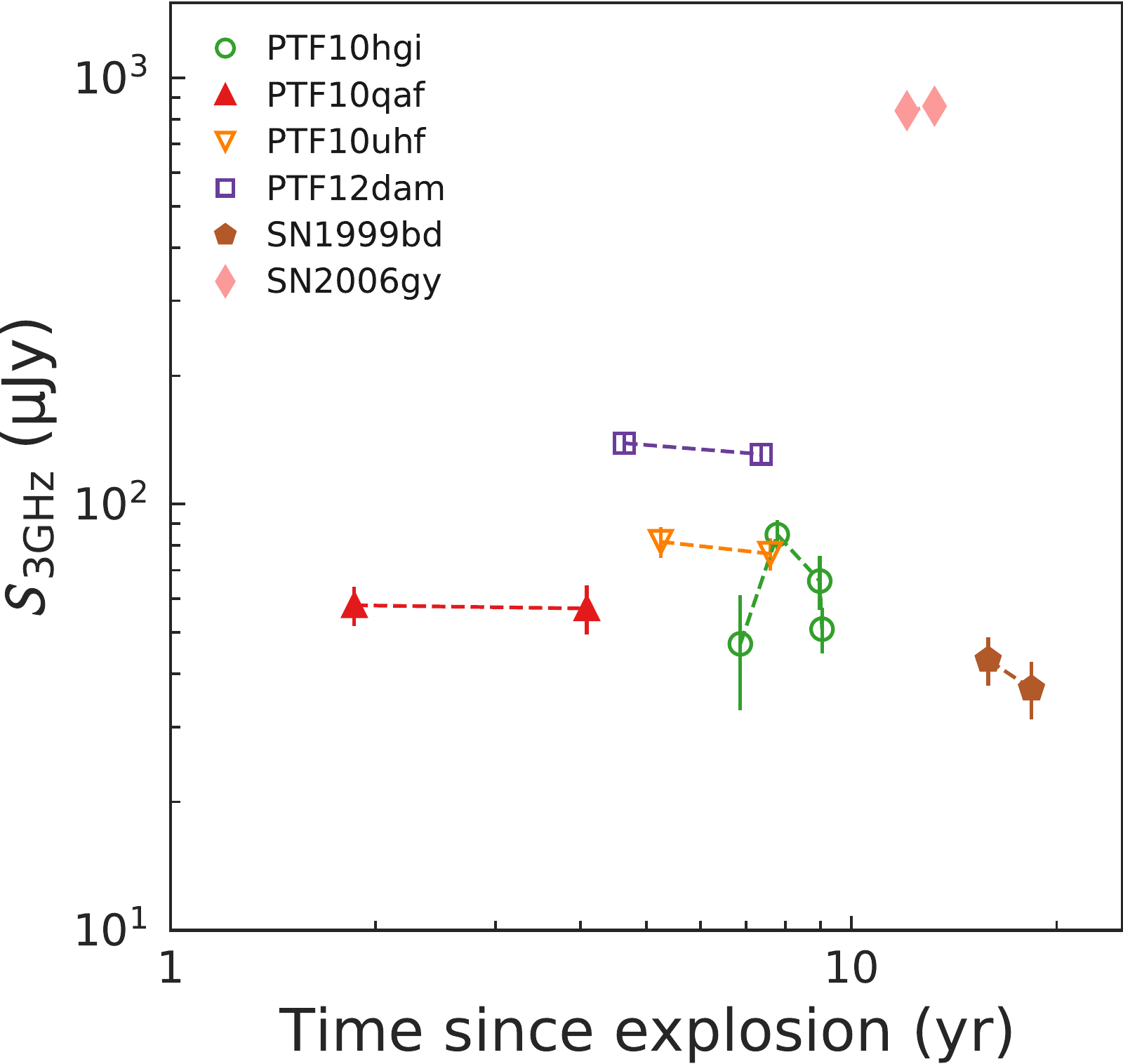}
\end{center}
\caption{
Observed 3 GHz flux densities as function of time since explosion (rest frame) for radio-detected sources. 
Open and filled symbols represent SLSNe-I and SLSNe-II, respectively. 
Data points of PTF10hgi include results of \cite{law19}, \cite{mond20}, and \cite{hats21}. 
\label{fig:lightcurve}}
\end{figure}

To see whether the sources show a time variability, we plot the observed flux densities as a function of time in Figure~\ref{fig:lightcurve} including previous 3 GHz observations by \cite{efte19} and \cite{mond20} for PTF10hgi. 
A significant variability is found in PTF10hgi, which was reported by \cite{hats21}. 
The flux densities of SN\,1999bd appear to be decreasing with time but are still consistent within flux uncertainties. 
The other sources do not show a significant variability in the timescale of years, suggesting the emission arise from activity in the hosts. 
This is supported by the fact that the emission of PTF10qaf, PTF10uhf, and SN\,2006gy are dominated by the galaxy center rather than the SN positions.

In the following discussions, we adopt the 3 GHz flux densities for the six $>$5$\sigma$ sources and 3$\sigma$ upper limits for the remaining sources. 
We consider that the origin of the radio emission is likely to be activities in the hosts for PTF10qaf, PTF10uhf, PTF12dam, SN\,1999bd, and SN\,2006gy, and the SN and/or the host for PTF10hgi \citep{hats21}.

\begin{deluxetable}{lc}
\tablecaption{Offset between radio peak and SN positions. \label{tab:offset}}
\tablewidth{0pt}
\tablehead{
\colhead{SLSN} & \colhead{Offset} \\
\colhead{} & \colhead{(arcsec)} 
}
\startdata
PTF10hgi   & 0.7 \\
PTF10qaf   & 4.0 \\
PTF10uhf   & 3.6 \\
PTF12dam   & 0.0 \\
SN\,1999bd & 3.0 \\
SN\,2006gy & 2.3 \\
\enddata
\end{deluxetable}

\section{Discussion} \label{sec:discussion}

Late-time radio observations provide useful implications for the obscured star formation in the host galaxies and power source of SLSNe. 
By using the radio data, we constrain radio emission arising from 
star-forming activities in the hosts (Section~\ref{sec:host}), 
off-axis afterglows (Section~\ref{sec:afterglow}),  
and pulsar/magnetar wind nebulae (Section~\ref{sec:magnetar}).

\subsection{Obscured Star Formation in Host Galaxies} \label{sec:host}

In the Baldwin-Phillips-Terlevich \citep[BPT;][]{bald81} diagram, the hosts of PTF09cnd, PTF10qaf, PTF12dam, and SN\,1999bd lie at the location of star-forming galaxies \citep{lelo15, perl16}. 
Because the radio emission in PTF10hgi may arise from an AGN in the host or wind nebula powered by a magnetar \citep{hats21}, the flux density is treated as an upper limit for star-forming activity in the host. 
Assuming that the radio emission of detected hosts is dominated by star-forming activity, we derive SFRs by using the 3 GHz flux densities. 
SFRs are derived by adopting the calibraions of \cite{murp11, murp12, murp17}  normalized to the \cite{chab03} IMF as follows \citep{jime21}: 
\begin{eqnarray}
{\rm SFR} = 4.87 \times 10^{-29} L_{\rm 1.4\ GHz}, \\
L_{\rm 1.4\ GHz} = \frac{4 \pi D_L^2}{(1 + z)^{1 + \alpha}} \left( \frac{1.4}{\nu_{\rm obs}} \right)^{\alpha} S_{\rm obs},
\end{eqnarray}
where $L_{\rm 1.4\ GHz}$ is the 1.4 GHz luminosity in erg~s$^{-1}$~Hz$^{-1}$, $D_L$ is the luminosity distance in cm, 
$\alpha$ is the spectral index (defined as $S_{\nu} \propto \nu^{\alpha}$), $\nu_{\rm obs}$ is the observed frequency in GHz, and $S_{\rm obs}$ is the observed flux density in erg~s$^{-1}$~cm$^{-2}$~Hz$^{-1}$. 
If a source was observed twice, then the flux density of combined analysis (Table \ref{tab:observations}) is used. 
The spectral index $\alpha$ is known to lie between around $-0.8$ and $-0.7$ for star-forming galaxies \citep[e.g.,][]{cond92}. 
\cite{smol17a} and \cite{delh17} found that the median spectral index between 1.4 and 3 GHz for $z < 2$ star-forming galaxies is consistent with $\alpha^{\rm 3 GHz}_{\rm 1.4 GHz} = -0.7$ with a standard deviation of $\sigma = 0.35$ based on the VLA-COSMOS 3 GHz data taking into account the lower limits for sources undetected at 1.4 GHz. 
In this study we assume $\alpha = -0.7$ following these studies. 
The derived 1.4 GHz luminosities and SFRs are listed in Table~\ref{tab:sfr}. 
Note that the SFRs would increase by about 10\% if we assume $\alpha = -0.8$.

We employed a stacking analysis on the 3 GHz maps of the undetected sources to obtain ``averaged'' properties of the hosts. 
No significant emission was detected for the samples of all 15 nondetections, 11 SLSNe-I, and 4 SLSNe-II. 
The obtained rms noise levels were 1.4, 1.8, and 2.3 $\mu$Jy~beam$^{-1}$ for the samples of all nondetections, SLSNe-I, and SLSNe-II , respectively. 
In order to constrain obscured star formation, we derived SFRs from stacking analysis. 
Because the sample of nondetections covers a wide redshift range ($z = 0.074$--0.287), where luminosity distances largely differ among the samples, we divided the sample into two groups:  6 SLSNe with $z = 0.074$--0.143 ($z_{\rm median} = 0.12$) and 9 SLSNe with $z = 0.205$--0.287 ($z_{\rm median} = 0.26$). 
The stacked images show no significant emission with rms noise levels of 2.2 and 1.8 $\mu$Jy~beam$^{-1}$, respectively. 
The corresponding 3$\sigma$ upper limits on SFRs are $<$0.20 and $<$0.88 $M_{\odot}$~yr$^{-1}$, respectively, which were calculated in the same manner as for individual hosts assuming the median redshifts (Table~\ref{tab:stacking}).

\begin{figure}
\begin{center}
\includegraphics[width=\linewidth]{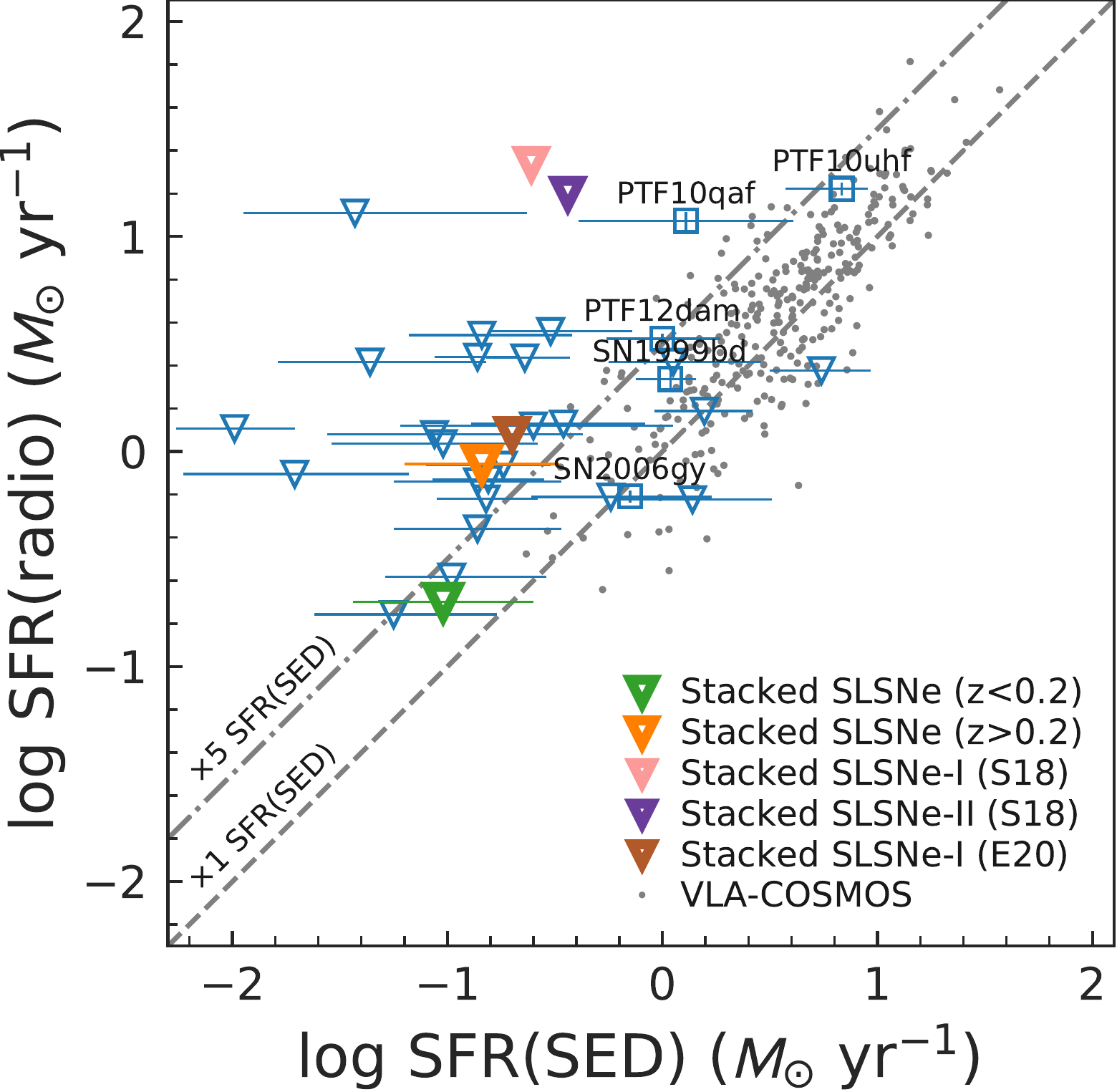}
\end{center}
\caption{
Comparison of SFRs derived from radio observations and SED analysis. 
Results on individual host and stacked results for radio-undetected SLSNe hosts are presented. 
SED-based SFRs are taken from \cite{schu18} except for SN\,1999bd and SN\,2006gy, 
where H$\alpha$-based SFR of \cite{lelo15a} and IR-based SFR of \cite{smit07} are adopted. 
Triangles represent 3$\sigma$ upper limits for nondetections. 
Stacked results on sample of nondetections (6 SLSNe at $z < 0.2$ as a green triangle and 9 SLSNe at $z > 0.2$ as an orange triangle) are plotted, where horizontal value and range are median and standard deviation of SFRs for sample \citep{schu18}, respectively. 
We also plot stacked results by \cite{schu18} (17 SLSNe-I and 13 SLSNe-II at $z < 0.5$) and \cite{efte21} (13 SLSNe-I at $z < 0.4$). 
For comparison, star-forming galaxies without AGN feature from VLA-COSMOS survey source catalog \citep{smol17a, smol17} at $z < 0.3$ (dots). 
Dashed and dot-dashed lines represent SFR(radio) equals to and five times SFR(SED), respectively. 
}
\label{fig:sfr}
\end{figure}

In Figure~\ref{fig:sfr}, we compare the SFRs derived from the radio observations with those from the SED modeling based on multi-wavelength data from rest-frame UV to NIR by \cite{schu18}. 
Because of the complexity of star-formation histories of the hosts of SN\,1999bd and SN\,2006gy, \cite{schu18} adopted the H$\alpha$ of \cite{lelo15a} and the IR luminosity of \cite{smit07} as an SFR indicator instead of using the results of SED analysis. 
The hosts of PTF10qaf, PTF10uhf, PTF12dam, and SN\,1999bd have an excess of radio-based SFRs over SED-based SFRs by a factor of 3--12. 
The radio excess in the hosts of SN\,1999bd, PTF10qaf, and PTF10uhf are consistent with the results by \citep{hats18}. 
On the other hand, \cite{hats18} and \cite{efte21} did not find a radio excess in the PTF12dam host. 
This is due to the difference in comparison SFRs adopted by \cite{hats18} and \cite{efte21}, where they used the results of \cite{perl16}. 
The H$\alpha$-based SFR of 3--5 $M_{\odot}$~yr$^{-1}$ \citep{lelo15, thon15, chen15, perl16} is consistent with our radio SFR of $3.4 \pm 0.2$ $M_{\odot}$~yr$^{-1}$. 
The same is true of PTF10uhf, where the radio SFR of $17 \pm 1$ $M_{\odot}$~yr$^{-1}$ is consistent with the extinction-corrected H$\alpha$ SFR of $19.36^{+7.301}_{-5.764}$ $M_{\odot}$~yr$^{-1}$ \citep{perl16}. 
\cite{schu18} noted that a systematic uncertainty of 0.3 dex is expected in the SED-based SFRs. 
Therefore, the existence of obscured star formation in the hosts of PTF10uhf and PTF12dam is inconclusive. 
Optical observations show that a dwarf galaxy hosting PTF10qaf is likely to be interacting with a large spiral galaxy \citep{perl16}. 
The optical image of the PTF10uhf host is also likely to represent a merger \citep{perl16}. 
The interaction or merger may cause dusty star formation in the hosts. 
The stellar masses of the hosts of SN\,1999bd, SN\,2006gy, PTF10qaf, and PTF10uhf are in the largest range among the sample of SLSNe \citep{lelo15, perl16, schu18}. 
In particular, the hosts of SN\,2006gy and PTF10uhf have the largest stellar mass with $\log{M_*/M_{\odot}} > 11$ \citep{perl16, schu18}. 
It is known that larger stellar mass galaxies have higher fraction of obscured star formation \citep[e.g.,][]{whit17}. 
\cite{whit17} found that more than 90\% of star formation is obscured for galaxies with $\log{M_*/M_{\odot}} > 10.5$. 
Although the host of SN\,2006gy, NGC~1260, is a S0/Sa galaxy, it shows signatures of current star formation with a far-IR luminosity of $\log{L_{\rm FIR}/L_{\odot}} = 9.85$ \citep{meus00, smit07} 
The IR-based SFR is 0.9 $M_{\odot}$~yr$^{-1}$, which is comparable to the radio-based SFR.

For the other SLSN hosts, we do not find a significant excess of the radio-based SFRs over the SED-based SFRs, although many of them have upper limits. 
The fact that there is no significant dust-obscured star formation in SLSN hosts is consistent with the radio observations by \cite{schu18} and \cite{efte21} for individual galaxies or stacked results.

\begin{figure}
\begin{center}
\includegraphics[width=\linewidth]{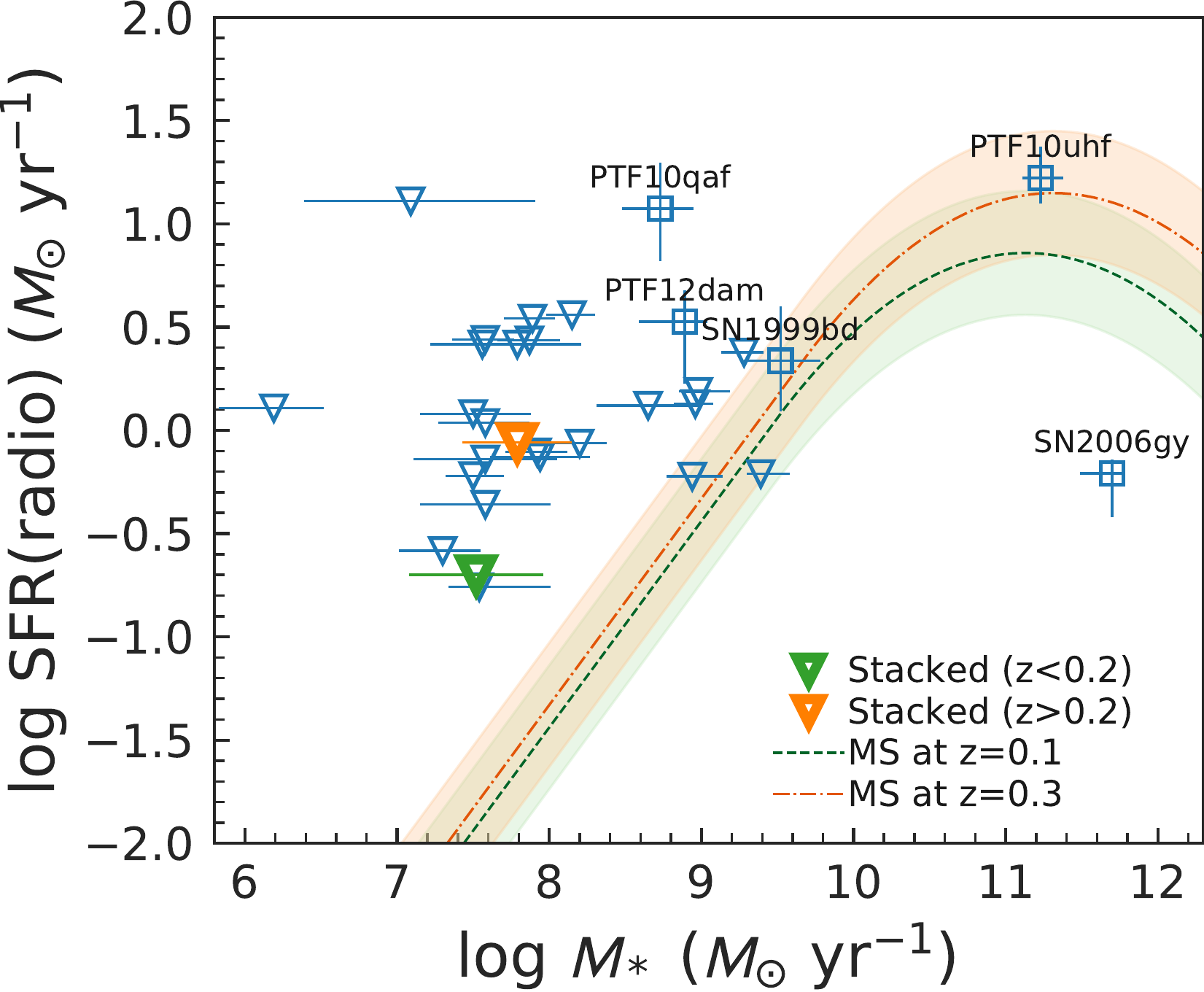}
\end{center}
\caption{
Radio-based SFRs as a function of stellar mass. 
Results on individual host and stacked results for radio-undetected hosts are presented. 
Horizontal value and range for stacked results are median and standard deviation for sample, respectively. 
Triangles represent 3$\sigma$ upper limits. 
Dashed and dot-dashed curves show main-sequence of star-forming galaxies at $z = 0.1$ and 0.3, respectively, with $\pm$0.3 dex uncertainty \citep{schr15}. 
}
\label{fig:ms-sfr}
\end{figure}

In Figure~\ref{fig:ms-sfr}, we plot radio-based SFRs as a function of stellar mass. 
It is known that star-forming galaxies follow a tight correlation between stellar mass and SFR, referred to as a galaxy main sequence.
The hosts of PTF10qaf and PTF12dam are located above the main sequence, suggesting that they have a starburst nature \citep[e.g.,][]{rodi11, elba11}. 
We note that if we adopt the stellar mass of $\log{M_*/M_{\odot}} = 10.24^{+0.22}_{-0.17}$ for PTF10qaf derived from the SED analysis by \cite{lelo15}, the host is on the main sequence. 
The high sSFR and starburst nature of the PTF12dam host are consistent with the results of optical spectroscopic observations \citep{thon15, chen15}. 
The SN\,2006gy host is below the main sequence, suggesting a quenched property. 
This is consistent with its classification of S0/Sa, but at the same time the host has ongoing star formation as found in this work and previous studies \citep{smit07}. 
The other radio-detected hosts of SN\,1999bd and PTF10uhf are within the range of the main sequence. 
The hosts of SLSNe are argued to have higher sSFR than main-sequence galaxies \citep[e.g.,][]{lelo15, schu18}. 
Our results are consistent with these previous studies overall but are not stringent because of the limited sensitivity of the radio observations.

\begin{deluxetable}{ccCC}
\tablecaption{1.4 GHz luminosities and radio-based SFRs of the hosts\label{tab:sfr}}
\tablewidth{0pt}
\tablehead{
\colhead{SLSN} & \colhead{$z$} & \colhead{$L_{\rm 1.4 GHz}$\tablenotemark{a}} & \colhead{SFR} \\
\colhead{} & \colhead{} & \colhead{(erg~s$^{-1}$~Hz$^{-1}$)} & \colhead{($M_{\odot}$~yr$^{-1}$)}
}
\startdata
CSS121015  & 0.286  & <7.4 \times 10^{28} & <3.6 \\
LSQ12dlf   & 0.255  & <5.4 \times 10^{28} & <2.6 \\
LSQ14mo    & 0.2561 & <7.2 \times 10^{28} & <3.5 \\
PTF09cnd   & 0.258  & <5.6 \times 10^{28} & <2.5 \\
PTF10hgi   & 0.0987 & <2.2 \times 10^{28} & <1.1 \\
PTF10qaf   & 0.284  & (2.4\pm0.2) \times 10^{29} & 12\pm1 \\
PTF10uhf   & 0.2882 & (3.4\pm0.2) \times 10^{29} & 17\pm1 \\
PTF12dam   & 0.107  & (6.9\pm0.4) \times 10^{28} & 3.4\pm0.2 \\
SN\,1999as & 0.127  & <1.2 \times 10^{28} & <0.6 \\
SN\,1999bd & 0.151  & (4.5\pm0.5) \times 10^{28} & 2.2\pm0.2 \\
SN\,2006gy & 0.019  & (1.3\pm0.1) \times 10^{28} & 0.62\pm0.04 \\
SN\,2006tf & 0.074  & <3.6 \times 10^{27} & <0.17 \\
SN\,2007bi & 0.128  & <1.6 \times 10^{28} & <0.79 \\
SN\,2007bw & 0.140  & <1.3 \times 10^{28} & <0.62 \\
SN\,2008es & 0.205  & <2.6 \times 10^{28} & <1.3 \\
SN\,2009nm & 0.210  & <2.7 \times 10^{28} & <1.3 \\
SN\,2010kd & 0.101  & <5.4 \times 10^{27} & <0.26 \\
SN\,2011ep & 0.280  & <5.4 \times 10^{28} & <2.6 \\
SN\,2011ke & 0.143  & <1.2 \times 10^{28} & <0.60 \\
SN\,2011kf & 0.245  & <5.7 \times 10^{28} & <2.8 \\
SN\,2012il & 0.175  & <1.8 \times 10^{28} & <0.87 \\
SN\,2013dg & 0.265  & <2.7 \times 10^{29} & <13 \\
SN\,2015bn & 0.110  & <2.5 \times 10^{28} & <1.3 \\
\enddata
\tablecomments{
Errors only take into account flux measurement uncertainty (1$\sigma$).
Limits are 3$\sigma$ (except for PTF10hgi, where radio emission observed in April 2020 was used as an upper limit.}
\tablenotemark{a}{Assuming a spectral index of $\alpha^{\rm 3 GHz}_{\rm 1.4 GHz} = -0.7$.}
\end{deluxetable}

\begin{deluxetable}{CcccC}
\tablecaption{Results for stacking analysis of radio-undetected SLSN hosts \label{tab:stacking}}
\tablewidth{0pt}
\tablehead{
\colhead{Sample} & \colhead{Number} & \colhead{$z_{\rm med}$} &\colhead{RMS} & \colhead{SFR\tablenotemark{a}} \\
\colhead{} & \colhead{} & \colhead{} & \colhead{($\mu$Jy~beam$^{-1}$)} & \colhead{($M_{\odot}$~yr$^{-1}$)}
}
\startdata
z < 0.2 & 6 & 0.12 & 2.2 & <0.20 \\
z > 0.2 & 9 & 0.26 & 1.8 & <0.88 \\
\enddata
\tablenotemark{a}{3$\sigma$ upper limits on SFRs assuming spectral index of $\alpha^{\rm 3 GHz}_{\rm 1.4 GHz} = -0.7$ and median redshift $z_{\rm med}$.}
\end{deluxetable}

\subsection{Afterglows from Off-axis Jet} \label{sec:afterglow}

A shock interaction between SN ejecta and CSM produces synchrotron radio emission. 
It is possible that radio afterglows arise from initially off-axis jets that decelerated and spread into the line of sight at late times. 
\cite{copp18} compiled all radio observations of SLSNe-I (9 SLSNe-I at that time) and constrained energies and mass-loss rates or CSM densities for an off-axis jet model. 
\citet{efte21} generated a grid of afterglow models for a range of jet energies and CSM densities for the sample of 15 SLSNe-I and compared them with the upper limits at 6 and 100 GHz. 
They ruled out the presence of jets with an isotropic-equivalent energy $E_{\rm iso} \gtrsim 10^{54}$ erg and CSM densities $n \sim 10^{-3}$--$10^2$ cm$^{-3}$ for observing angles $\theta_{\rm obs} = 30$ and 60$^{\circ}$.

\begin{figure*}
\begin{center}
\includegraphics[width=.9\linewidth]{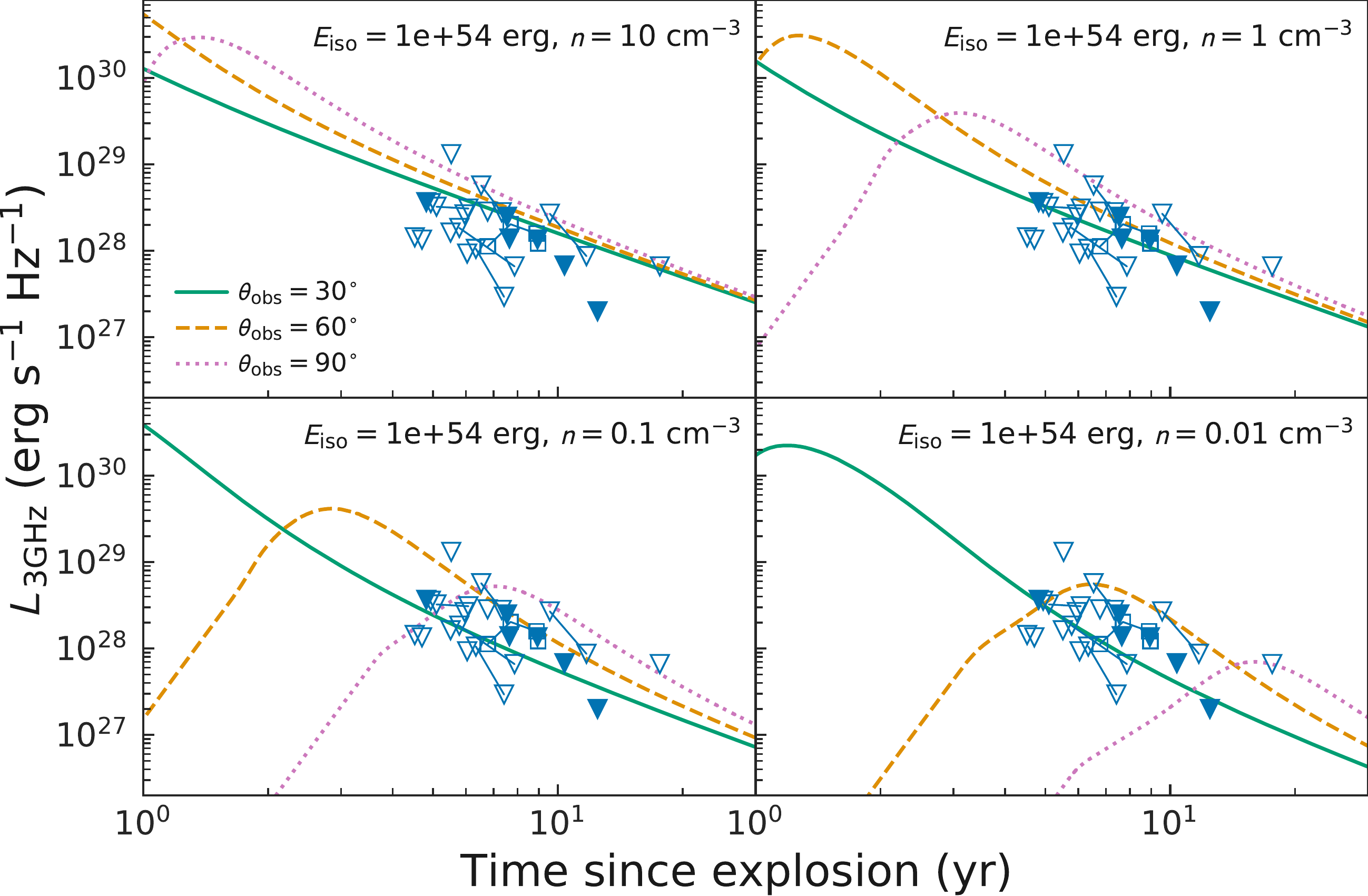}
\end{center}
\caption{
3 GHz luminosities of SLSNe as a function of time since explosion (rest frame) in comparison with afterglow models generated using the {\tt afterglowpy} code \citep{ryan20} assuming a median redshift of the sample of $z = 0.19$ with $E_{\rm iso} = 10^{54}$ erg and 
$n = 10$ cm$^{-3}$ (top left), 
$n = 1$ cm$^{-3}$ (top right), 
$n = 0.1$ cm$^{-3}$ (bottom left), 
and $n = 0.01$ cm$^{-3}$ (bottom right). 
Solid, dashed, and dotted curves represent light curves with viewing angles of $\theta_{\rm obs} = 30^{\circ}$, $60^{\circ}$, and $90^{\circ}$, respectively. 
Triangles represent 3$\sigma$ upper limits. 
Open and filled symbols represent SLSNe-I and SLSNe-II, respectively. 
SLSNe whose hosts are radio-detected are not plotted.
Same source with different observing epochs is connected with a solid line. 
}
\label{fig:afterglow_time-lum}
\end{figure*}

\begin{figure*}
\begin{center}
\includegraphics[width=.32\linewidth]{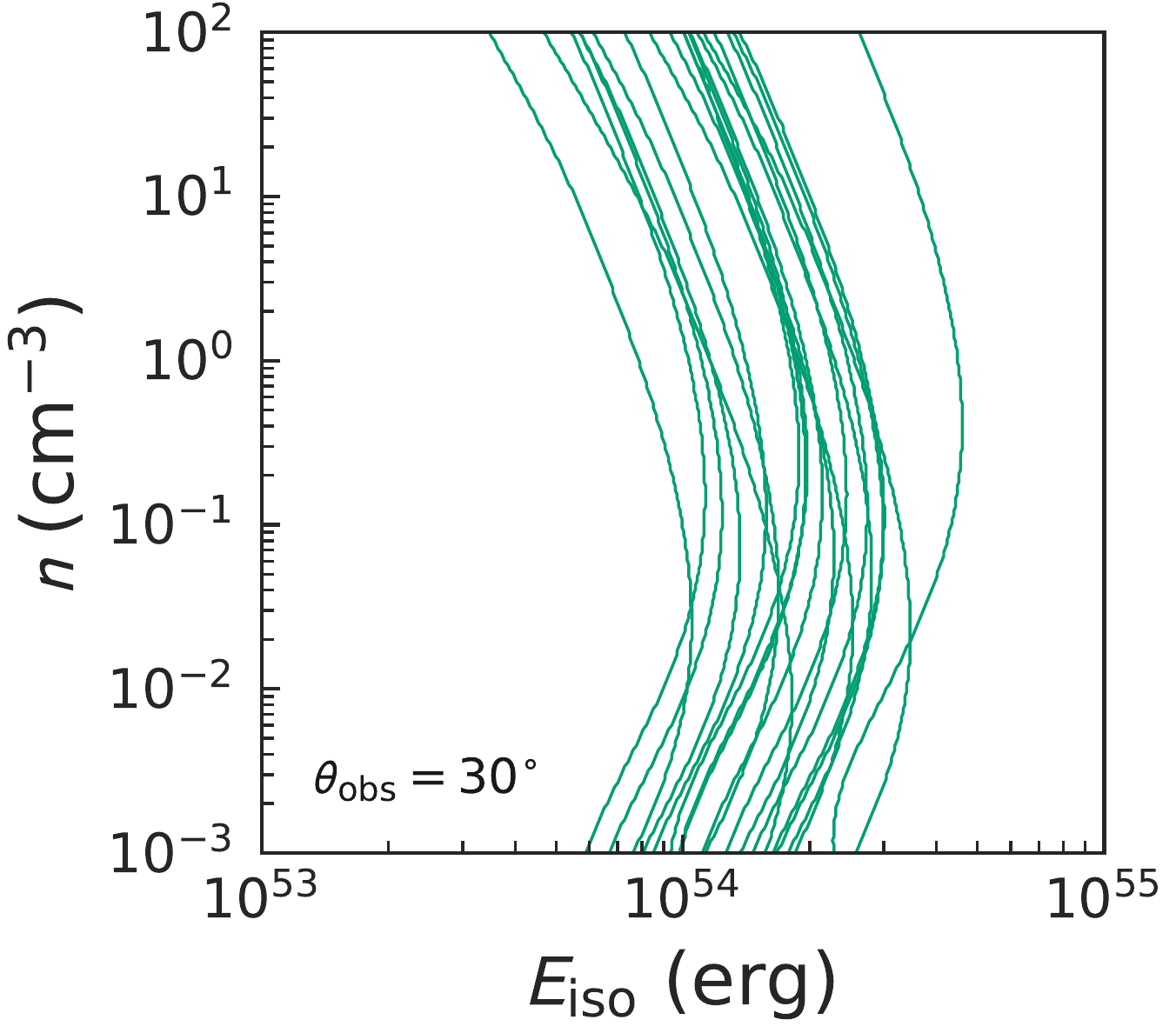}
\includegraphics[width=.32\linewidth]{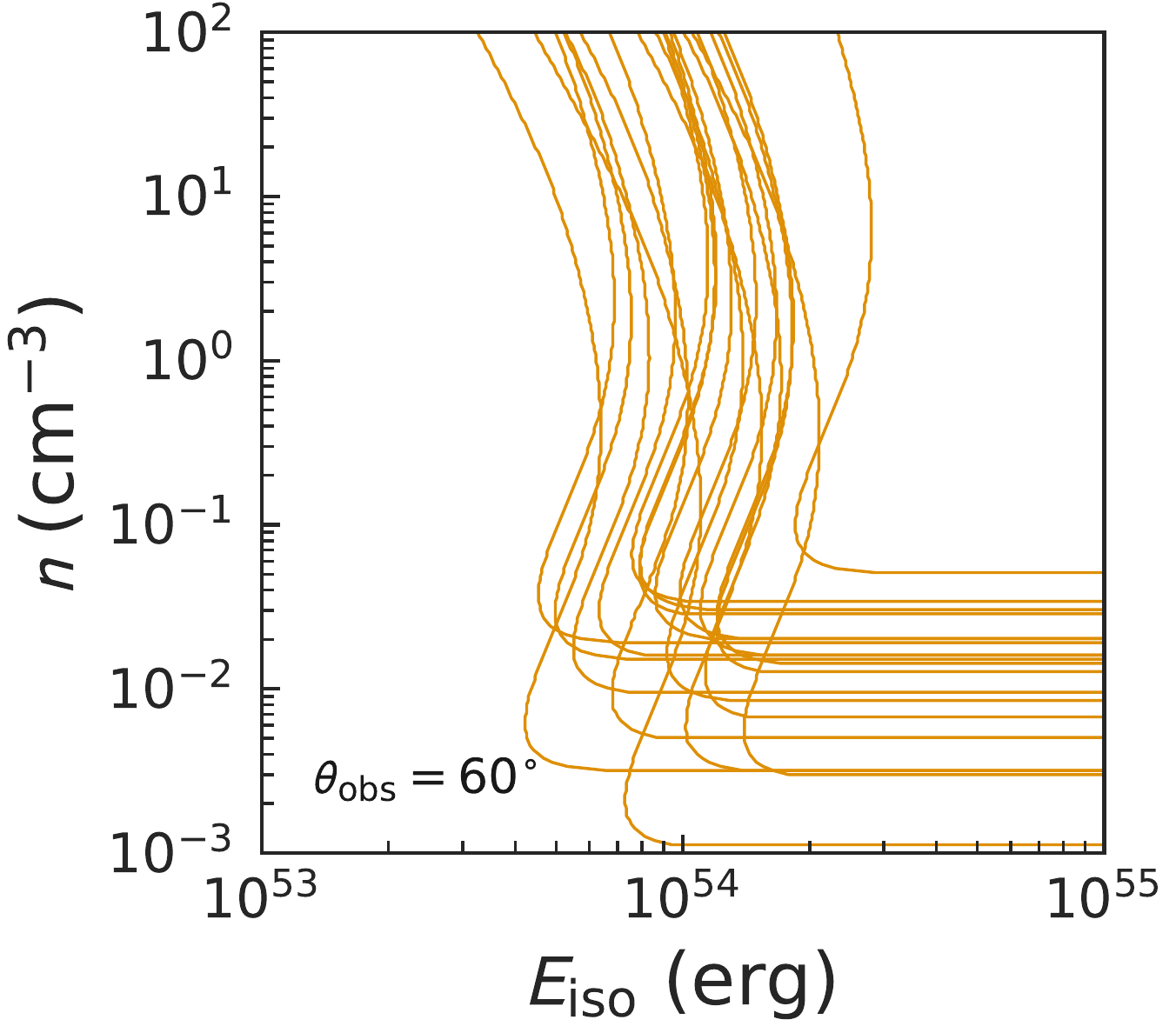}
\includegraphics[width=.32\linewidth]{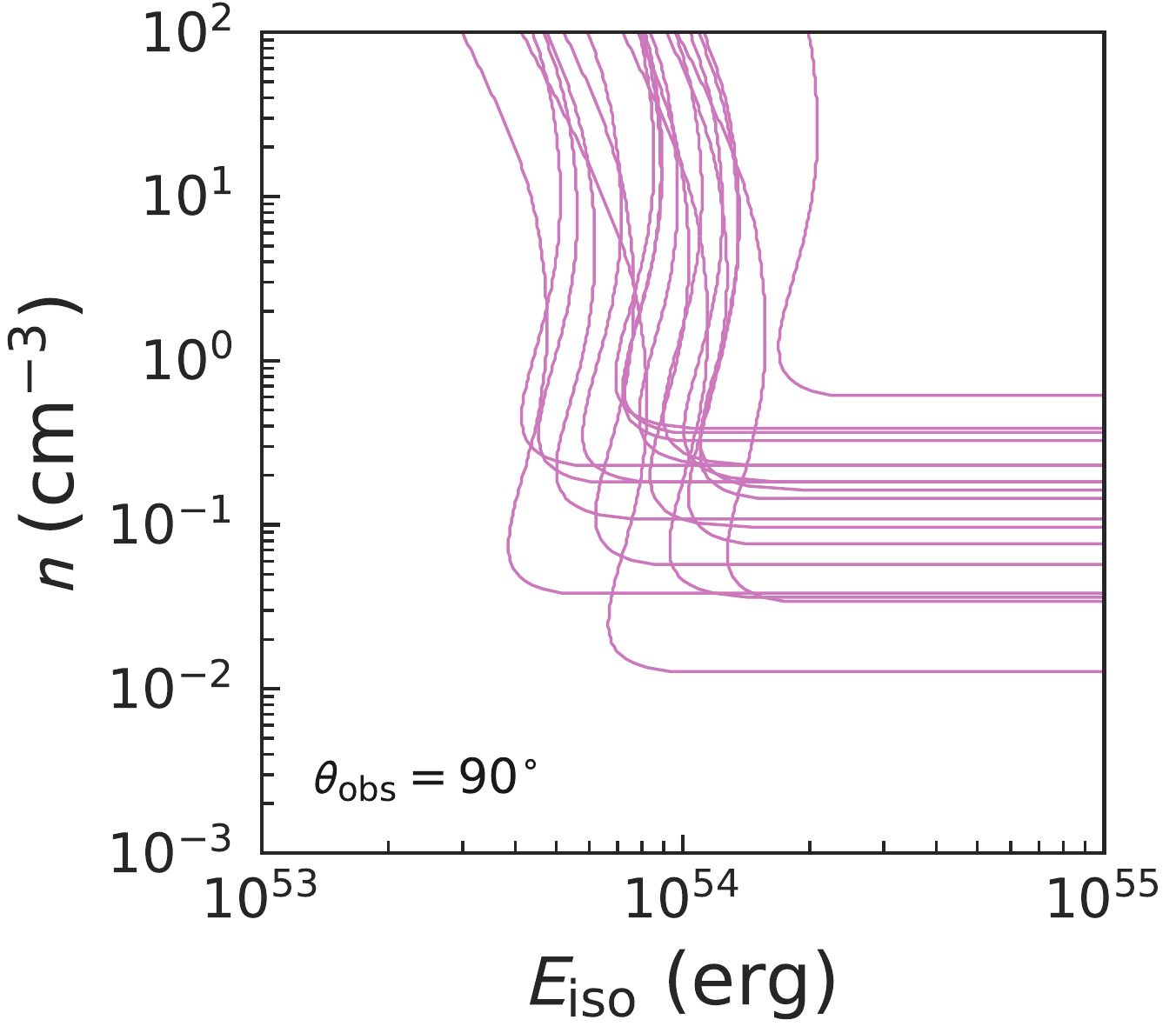}
\end{center}
\caption{
Pairs of isotropic equivalent energies ($E_{\rm iso}$) and CSM densities ($n$) which reach the observed 3 GHz flux density (or upper upper limit) for each SLSN calculated with the {\tt afterglowpy} code \citep{ryan20} with viewing angles of $\theta_{\rm obs} = 30^{\circ}$, $60^{\circ}$, and $90^{\circ}$. 
A lower left (lower energy and density) region of each line represents the allow parameters for each SLSN.
}
\label{fig:n-Eiso}
\end{figure*}

By using our 3 GHz radio upper limits, we constrain the parameters for an afterglow model. 
We utilize the publicly available {\tt afterglowpy} code \citep{ryan20}, which generates afterglow light curves using semianalytic approximations of the jet evolution and synchrotron emission. 
The {\tt afterglowpy} code has been calibrated to the {\tt BoxFit} code \citep{van12} and produces similar light curves for on- and off-axis top hat jets. 
We adopted a top hat jet with following fixed parameters: 
jet opening angle of $10^{\circ}$, 
electron energy distribution index of $p = 2.5$, 
thermal energy fraction in electrons of $\epsilon_e = 0.1$, 
and thermal energy fraction in magnetic field of $\epsilon_B = 0.01$, which are typical values for GRBs \citep[e.g.,][]{wang15}; these values were also assumed in previous studies \citep{efte19, efte21}. 
We generated light curves with 
isotropic equivalent energies ranging from $10^{53}$ to $10^{55}$ erg,
CSM densities ranging from $10^{-3}$ to 10$^2$ cm$^{-3}$, 
and viewing angles of 30, 60, and 90$^{\circ}$. 
Figure~\ref{fig:afterglow_time-lum} compares the VLA 3 GHz data with the model light curves with $E_{\rm iso} = 10^{54}$ erg and $n =$ 10, 1, 0.1, and 0.01 cm$^{-3}$. 
Figure~\ref{fig:n-Eiso} shows allowed parameters in $n-E_{\rm iso}$ plots estimated for individual SLSN. 
We found that the radio upper limits exclude the models with 
i) $E_{\rm iso} \gtrsim$ several $\times$ $10^{53}$ erg and $n \gtrsim 10^{-3}$ cm$^{-3}$ for $\theta_{\rm obs} \sim 30$--$60^{\circ}$ and
ii) $E_{\rm iso} \gtrsim$ several $\times$ $10^{53}$ erg and $n \gtrsim 10^{-2}$ cm$^{-3}$ for $\theta_{\rm obs} \sim 90^{\circ}$.
Afterglows with lower energies or lower CSM densities are not excluded with the current data. 
The results are consistent with previous studies overall \citep{nich16a, copp18, efte21}. 
Further observations are required to constrain the afterglow models or parameters.

\subsection{Pulsar/Magnetar Wind Nebulae} \label{sec:magnetar}

It is predicted that late-time radio emission may arise from PWNe caused by pulsar/magnetar-driven SLSNe-I \citep[e.g.,][]{metz14, mura16, kash17, metz17}. 
Based on the model of \citet{mura15} and \citet{mura16}, \cite{oman18} and \cite{law19} calculated radio light curves for samples of SLSNe-I with maximum absorption and with no absorption processes in the PWN and SN ejecta. 
The initial parameters of a magnetar (spin period $P_i$, magnetic field $B$, and ejecta mass $M_{\rm ej}$) were obtained by fitting the early optical light curves by eye.
They defined two parameter sets for reproducing the optical light curves with $P = 1$ ms ($P_{\rm min}$) and with the largest spin period ($P_{\rm max}$). 
The radio emission can be absorbed in the PWN and the SN ejecta, but the system can be transparent in the timescale of years and reaches its peak at $\sim$10--30 yr after the explosions. 
\cite{efte21} considered two scenarios for PWNe: an ion-electron wind based on the prescription for FRB 121102 from \cite{marg18}, and an electron-positron wind following the methods of \cite{mura16}, \cite{oman18}, and \cite{mura21}. 
The inferred magnetar parameters (ejecta mass and velocity, magnetar magnetic field and spin period) are derived via Markov chain Monte Carlo fits to the multi-color light curves using {\tt MOSFiT}.

Our sample of SLSNe-I (except for PTF10uhf, SN\,1999as, and SN\,2011ep) are covered by these studies \citep{oman18, law19, efte21}, and we compare our VLA results with their predictions based on electron-positron wind models. 
The 3 GHz upper limits on SN\,2011ke and SN\,2012il exclude the predictions by \cite{oman18} and \cite{law19} with or without absorption processes. 
The predictions without absorption for PTF09cnd by \cite{law19} and \cite{efte21} and for SN\,2010kd by \cite{law19} are also excluded. 
While the predictions for SN\,2007bi by \cite{efte21} both with or without absorption are excluded, the $P_{\rm min}$ model with absorption by \cite{law19} is still viable. 
The predictions for the remaining SLSNe-I cannot be ruled out in the current data. 
The outcomes are summarized in Table~\ref{tab:model}.

Finally, we constrain the pulsar/magnetar-driven model for SN\,2015bn by \cite{mura21}. 
They found that the ALMA observations disfavors a PWN model motivated by the Crab nebula and argued that this tension can be resolved if the nebular magnetization is very high or very low. 
We found that at least the low-magnetization model without ejecta absorption is ruled out by our VLA 3 GHz upper limit.

\begin{deluxetable}{ccccc}
\tablecaption{Score sheet for magnetar wind nebular models constrained by this study \label{tab:model}}
\tablewidth{0pt}
\tablehead{
\colhead{SLSN} & \multicolumn{2}{c}{with absorption} & \multicolumn{2}{c}{without absorption} 
}
\startdata
 & \multicolumn{4}{c}{Model of \cite{oman18}} \\
 & $P_{\rm min}$ & $P_{\rm max}$ & $P_{\rm min}$ & $P_{\rm max}$ \\
\tableline
SN\,2011ke & Excluded & Excluded & Excluded & Excluded \\
SN\,2012il & Excluded & Excluded & Excluded & Excluded \\
\tableline
 & \multicolumn{4}{c}{Model of \cite{law19}} \\
 & $P_{\rm min}$ & $P_{\rm max}$ & $P_{\rm min}$ & $P_{\rm max}$ \\
\tableline
PTF09cnd   &          & Excluded &          & Excluded \\
SN\,2007bi &          & Excluded & Excluded & Excluded \\
SN\,2010kd &          & Excluded &          & Excluded \\
SN\,2011ke & Excluded & Excluded & Excluded & Excluded \\
\tableline
 & \multicolumn{4}{c}{Model of \cite{efte21}} \\
\tableline
PTF09cnd   & \multicolumn{2}{c}{} & \multicolumn{2}{c}{Excluded} \\
SN\,2007bi & \multicolumn{2}{c}{Excluded} & \multicolumn{2}{c}{Excluded} \\
\enddata
\end{deluxetable}

\section{Conclusions} \label{sec:conclusions}

We reported the results of VLA 3 GHz continuum observations of 23 SLSNe and their host galaxies 5--21 yr after the explosions. 
The sample consisted of 15 SLSNe-I and 8 SLSNe-II at $z < 0.3$. 
This study provided one of the largest samples of SLSNe with late-time radio data. 
We detected radio emission in six sources (PTF10hgi, PTF10qaf, PTF10uhf, PTF12dam, SN\,1999bd, SN\,2006gy) with S/N $>$ 5, and tentatively detected in two sources (SN\,2007bw and SN\,2012il) with S/N $\sim$ 3.6. 
The key findings are as follows: 
\begin{enumerate}
\item 
The radio-detected sources were observed more than once, and we examined the time variability. 
No significant variability was found in the $>$5$\sigma$ sources except for PTF10hgi, which was reported to have a variability in the literature. 
This suggests that the emission arise from activity in the hosts for PTF10qaf, PTF10uhf, PTF12dam, SN\,1999bd, and SN\,2006gy, coupled with the fact that the radio peak positions are consistent with the galaxy centers. 

\item 
We compared SFRs derived from the 3 GHz flux densities with SFRs from the SED modelling based on rest-frame UV to NIR data. 
We found that four hosts (PTF10qaf, PTF10uhf, PTF12dam, and SN\,1999bd) have an excess of radio-based SFRs over SED-based SFRs, suggesting that there exists obscured star formation that cannot be traced by UV--NIR data. 
The obscured star formation is consistent with their starburst nature and/or large stellar masses. 
The upper limits for the undetected hosts and stacked results show that the majority of the SLSN hosts do not have a significant obscured star formation.

\item 
We compared radio-based SFRs and stellar masses and found that the hosts of PTF10qaf and PTF12dam are located above the galaxy main sequence, 
the hosts of SN\,1999bd and PTF10uhf are within the range of the main sequence, 
and the SN\,2006gy host is below the main sequence. 
Our results are consistent overall with the fact that the hosts of SLSNe have higher sSFR compared to main-sequence galaxies reported in the literature, but are not stringent because of the limited sensitivity of the radio observations. 

\item 
By using our 3 GHz radio upper limits, we constrained the parameters for afterglows arising from off-axis jets. 
We found that the radio upper limits excluded the models with a higher energy ($E_{\rm iso} \gtrsim$ several $\times$ $10^{53}$ erg) and higher CSM densities ($n \gtrsim 0.01$ cm$^{-3}$), but lower energies or lower CSM densities were not excluded with the current data. 

\item 
We constrained the models of electron–positron PWNe caused by pulsar/magnetar-driven SLSNe by comparing the 3 GHz upper limits with the predictions in the literature. 
The models for some of the SLSNe in our sample (PTF09cnd, SN\,2007bi, SN\,2010kd, SN\,2011ke, SN\,2012il, and SN\,2015bn) were constrained, but the predictions for the remaining SLSNe-I were not ruled out in the current data. 
\end{enumerate}

So far, about 30 SLSNe were observed in radio and only a few SLSNe had radio detections. 
In addition, the current sensitivity is insufficient to constrain parameters of models for SLSNe or obscured star formation in their hosts. 
High-sensitivity, long-term monitoring observations of a large sample will allow us to examine the physical nature of SLSNe. 
Observations with future radio telescopes such as 
the Square Kilometer Array (SKA)\footnote{\url{https://www.skatelescope.org/}} and 
the Next Generation VLA (ngVLA)\footnote{\url{https://ngvla.nrao.edu/}} 
will open a new window toward understanding extreme transient events.

\vspace{\baselineskip}

We thank the referee for helpful comments and suggestions which significantly improved the paper. 
We are grateful to Masao Hayashi and the PDJ collaboration for fruitful discussions. 
We would like to acknowledge NRAO staffs for their help in preparation of observations. 
BH is supported by JSPS KAKENHI Grant Number 19K03925. 

The National Radio Astronomy Observatory is a facility of the National Science Foundation operated under cooperative agreement by Associated Universities, Inc.

Based on observations made with the NASA/ESA Hubble Space Telescope, and obtained from the Hubble Legacy Archive, which is a collaboration between the Space Telescope Science Institute (STScI/NASA), the Space Telescope European Coordinating Facility (ST-ECF/ESA) and the Canadian Astronomy Data Centre (CADC/NRC/CSA).

This project used public archival data from the Dark Energy Survey (DES). Funding for the DES Projects has been provided by the U.S. Department of Energy, the U.S. National Science Foundation, the Ministry of Science and Education of Spain, the Science and Technology FacilitiesCouncil of the United Kingdom, the Higher Education Funding Council for England, the National Center for Supercomputing Applications at the University of Illinois at Urbana-Champaign, the Kavli Institute of Cosmological Physics at the University of Chicago, the Center for Cosmology and Astro-Particle Physics at the Ohio State University, the Mitchell Institute for Fundamental Physics and Astronomy at Texas A\&M University, Financiadora de Estudos e Projetos, Funda{\c c}{\~a}o Carlos Chagas Filho de Amparo {\`a} Pesquisa do Estado do Rio de Janeiro, Conselho Nacional de Desenvolvimento Cient{\'i}fico e Tecnol{\'o}gico and the Minist{\'e}rio da Ci{\^e}ncia, Tecnologia e Inova{\c c}{\~a}o, the Deutsche Forschungsgemeinschaft, and the Collaborating Institutions in the Dark Energy Survey.

The Collaborating Institutions are Argonne National Laboratory, the University of California at Santa Cruz, the University of Cambridge, Centro de Investigaciones Energ{\'e}ticas, Medioambientales y Tecnol{\'o}gicas-Madrid, the University of Chicago, University College London, the DES-Brazil Consortium, the University of Edinburgh, the Eidgen{\"o}ssische Technische Hochschule (ETH) Z{\"u}rich,  Fermi National Accelerator Laboratory, the University of Illinois at Urbana-Champaign, the Institut de Ci{\`e}ncies de l'Espai (IEEC/CSIC), the Institut de F{\'i}sica d'Altes Energies, Lawrence Berkeley National Laboratory, the Ludwig-Maximilians Universit{\"a}t M{\"u}nchen and the associated Excellence Cluster Universe, the University of Michigan, the National Optical Astronomy Observatory, the University of Nottingham, The Ohio State University, the OzDES Membership Consortium, the University of Pennsylvania, the University of Portsmouth, SLAC National Accelerator Laboratory, Stanford University, the University of Sussex, and Texas A\&M University.

Based in part on observations at Cerro Tololo Inter-American Observatory, National Optical Astronomy Observatory, which is operated by the Association of Universities for Research in Astronomy (AURA) under a cooperative agreement with the National Science Foundation.


\facility{Karl G. Jansky Very Large Array}

\software{
CASA \citep{mcmu07}, 
{\tt afterglowpy} \citep{ryan20}, 
{\sc astropy} \citep{robi13, pric18}, 
{\sc matplotlib} \citep{hunt07}
}


\listofchanges
\end{document}